\documentclass[a4paper,fleqn,usenatbib]{mnras}

\usepackage[T1]{fontenc}
\usepackage{ae,aecompl}

\usepackage{graphicx}	% Including figure files
\usepackage{amsmath}	% Advanced maths commands
\usepackage{amssymb}	% Extra maths symbols

\newcommand\eg{{\it e.g.} }

\newcommand\ie{{\it i.e.} }
\newcommand\etc{{\it etc} }
\newcommand\Herschel{{\it Herschel } }

\title[A wide binary trigger for white dwarf pollution ]{A wide binary trigger for white dwarf pollution }

\author[A. Bonsor et al.]{
Amy Bonsor$^{1,2}$\thanks{E-mail: abonsor@ast.cam.ac.uk}
and Dimitri Veras$^{3}$
\\
$^{1}$Institute of Astronomy, University of Cambridge, Madingley Road, Cambridge, CB3 0HA, UK\\
$^{2}$School of Physics, H.H. Wills Physics Laboratory, University of Bristol, Tyndall Avenue, Bristol BS8 1TL, UK\\
$^{3}$Department of Physics, University of Warwick, Coventry CV4 7AL, UK
}

\date{Accepted XXX. Received YYY; in original form ZZZ}

\pubyear{2015}

\begin{document}
\label{firstpage}
\pagerange{\pageref{firstpage}--\pageref{lastpage}}
\maketitle

\begin{abstract}

Metal pollution in white dwarf atmospheres is likely to be a signature of remnant planetary systems.  Most explanations for this pollution predict a sharp decrease in the number of polluted systems with white dwarf cooling age.  Observations do not confirm this trend, and metal pollution in old (1-5 Gyr) white dwarfs is difficult to explain.  We propose an alternative, time-independent mechanism to produce the white dwarf pollution.  The orbit of a wide binary companion can be perturbed by Galactic tides, approaching close to the primary star for the {\it first} time after billions of years of evolution on the white dwarf branch.  We show that such a close approach perturbs a planetary system orbiting the white dwarf, scattering planetesimals onto star-grazing orbits, in a manner that could pollute the white dwarf's atmosphere.  Our estimates find that this mechanism is likely to contribute to metal pollution, alongside other mechanisms, in up to a few percent of an observed sample of white dwarfs with wide binary companions, independent of white dwarf age.  This age independence is the key difference between this wide binary mechanism and others mechanisms suggested in the literature to explain white dwarf pollution.  Current observational samples are not large enough to assess whether this mechanism makes a significant contribution to the population of polluted white dwarfs, for which better constraints on the wide binary population are required, such as those that will be obtained in the near future with Gaia.

\end{abstract}


\section{Introduction}
%{\bf just a rough outline}
\label{sec:intro}

Elements heavier than helium sink rapidly in the atmospheres of white dwarfs, where sinking timescales are on the order of days to weeks (DA white dwarfs) and $10^4-10^6$ years (DB white dwarfs) \citep{Koester06} or see Fig. 1 of \cite{Wyattstochastic}. Observations of emission lines from metallic species in the atmospheres of white dwarfs \citep[e.g.][]{Koester1997, Zuckerman03, Melis10}, therefore, suggest the recent accretion of the observed material.  Accretion from the interstellar medium was ruled out by \cite{Farihi10ism, ism93, Jura2006, Kilic2007, Barstow2014}.  These observations are thought to indicate the accretion of planetary material \citep[e.g.][]{Alcock86,DebesSigurdsson,JuraWD03, jurasmallasteroid, Kilic06}, in at least 25\% \citep{Zuckerman03, ZK10}, but maybe up to 50\% \citep{Koester2014} of white dwarfs.  Dusty and/or gaseous discs, very close ($r<R_\odot$) to a subset of these polluted white dwarfs may be indicative of the accreting material \citep[e.g][]{ Kilic06, Gaensicke06, vonHippel07, Jura07,Gaensicke2007}. The composition of the accreted material, when analysed in detail supports the planetary material hypothesis \citep[e.g.][]{Klein10, Klein2011, Gaensicke2012}.

Planetary systems are common on the main-sequence, and throughout the Milky Way they are the rule, not the exception \citep{Cassan2012}.  Observations with HARPS suggest that at least 50\% of solar-type stars harbour at least one planet with a period less than 100 days \citep{Mayor2011}, whilst \Herschel found that at least 20\% of FGK stars have a detectable debris disc \citep{Eiroa2013}. Although close-in ($<1-5$AU) planets may be swallowed by the expanding giant's envelope \citep{villaverlivio,Villaver09, Mustill_tidal_2012, Adams2013, Villaver2014}, there is no evidence that all outer planetary systems are destroyed \citep[e.g.][]{juraotherkb, bonsor10}. Dynamical instabilities in the outer planetary system, induced following stellar mass loss \citep[e.g][]{DebesSigurdsson, Veras11, Veras_Wyatt2012, Veras_tout_2012, Veras_twoplanet_2013, Voyatzis2013, Mustill_threeplanet, Veras2015_closein}, may scatter asteroids or comets onto star-grazing orbits \citep{bonsor11,bonsor_tiss, debesasteroidbelt, Frewen2014}, where they are tidally disrupted and accreted onto the star \citep[e.g.][]{Graham1990, JuraWD03, jurasmallasteroid, Bear2013,Veras_tidaldisruption1, Veras_tidaldisrupt2}.

Given a sufficiently massive planetesimal belt, it has been shown that such a theory can explain the observed metal pollution \citep{bonsor11, debesasteroidbelt, Frewen2014}. However, a steep decrease in the mass of asteroids/comets scattered onto star-grazing orbits with time after the formation of the white dwarf, is predicted \citep{bonsor11, debesasteroidbelt}. This finding would correspond to a steep decrease in the level of pollution (or fraction of systems with pollution), as a function of time.  Although late time instabilities may be induced in multi-planet systems \citep{Veras_twoplanet_2013, Mustill_threeplanet, Veras2015_closein}, the frequency of these will always be less than those around young white dwarfs.

Observationally there is no evidence that there are fewer, old polluted white dwarfs, nor that the level of pollution decreases with white dwarf cooling age \citep{Wyattstochastic, Koester2014}.  Even the archetypal polluted white dwarf, van Maanen's star \citep{vanMaanen1920}, has an effective temperature of 6,220K, or cooling age of $\sim3$Gyr \citep{wdcat09}, and there are many further observations of polluted, old, white dwarfs ($T_{\rm eff}<8,000$K, equivalent to cooling ages of up to 5 Gyrs \cite{farihi_g77-50,Koester11}).  Instabilities induced following stellar mass loss struggle to explain pollution for white dwarfs with cooling ages of Gyrs \citep{DebesSigurdsson, bonsor11, debesasteroidbelt}.  Alternative explanations for the pollution in these systems have been suggested, including stellar encounters \citep{farihi_g77-50}, or a relationship with magnetic fields \citep{Hollands2015}.  Others, such as exo-Oort cloud comet impacts \citep{VerasOort, Stone2015} and volatile sublimation of minor bodies in planet-less systems \citep{Veras2015}, have largely been ruled out.

Here, we suggest an alternative explanation. It has been shown for main-sequence planetary systems, that whilst in general wide ($a >1,000$AU) binary companions do not influence the dynamics of the planetary system, the orbit of the binary varies due to Galactic tides. During periods of `close'  pericentre passage, the binary may excite the eccentricities of planets, even ejecting them \citep{Kaib2013}. The increased eccentricity of exoplanets in systems with wide binary companions has been confirmed by observations \citep{Kaib2013}. Here, we propose that the same scenario could be applied to white dwarf planetary systems. If the white dwarf is orbited by a wide binary companion, the planetary system may remain unperturbed for billions of years, before Galactic tides alter the orbit of the companion such that it induces dynamical instabilities in the planetary system. These instabilities can lead to material being scattered onto star-grazing orbits and accreted onto the white dwarf, such that we observe pollution in the white dwarf's atmosphere.  This process has the advantage of being independent of the age of the white dwarf, although it depends critically on the population of wide binaries orbiting white dwarfs.

The population of wide ($>1,000$ AU) binaries is difficult to characterise, in general.  Common proper motions indicate that two stars may be related \citep[\eg][]{LepineBongiorno2007,Makarov2008}, but follow-up parallax and radial velocity observations are required to determine whether they are bound.  Wide binaries are, therefore, likely to be significantly more common than determined observationally.  Theoretical models predict a wide binary fraction of 1-30\%, with $10^3<a<0.1$pc \citep{Kouwenhoven2010}, whilst observations of solar-type stars \citep{LepineBongiorno2007, Raghavan2010} find a fraction of 9.5\% or 11.5\%, and observations of field white dwarfs yield a wide binary fraction of 22\% \citep{Farihi2005}.  There are, however, very few white dwarfs with known wide binary companions that have been searched for pollution, and many surveys for pollution have focussed on apparently single white dwarfs.  \cite{Zuckermanbinary} present a sample of 17 white dwarfs with companions separated by more than 1,000AU, searched for Ca II with Keck/VLT and find that 5 are polluted.  We note here that white dwarfs in close binaries may be polluted by interactions with the stellar wind of the companion, as discussed in \citep[\eg][]{Zuckerman03, Zuckermanbinary}, and for this reason we focus on `wide' binaries.  Fig. 8 of \cite{Veras_tout_2012} shows that the boundary between `close' and `wide' occurs at tens of AU.

In this paper, we illustrate the manner in which a wide binary companion could lead to pollution in a white dwarf.  This mechansim is likely to be just one of many that lead to pollution.  The critical difference of this mechanism is the absence of any dependence on white dwarf cooling age.  \S\ref{sec:galatic} shows how a wide binary's orbit can be altered by Galactic tides, such that a close approach occurs between the secondary and a planetary system orbiting the primary (white dwarf).  \S\ref{sec:pollution} presents simulations that show how this close approach can lead to planetesimals scattered onto star-grazing orbits, a pre-requisite for pollution.  In \S\ref{sec:fraction} we estimate the fraction of white dwarfs with wide binary companions where pollution might occur, which is compared to observations in \S\ref{sec:observations}. In \S\ref{sec:discussion} and \S\ref{sec:conclusions} we discuss our results and present our conclusions.

\section{The change in orbit of a wide binary due to the Galactic tide}
\label{sec:galatic}
We firstly illustrate the manner in which the Galactic tide can change the orbit of a wide binary. Above a critical separation, the Galactic field strength is different for the two components of the binary, such that their orbit evolves. Analytically the effect of the Galactic tide on the binary's orbit can be understood through the equations of motion for the binary's orbital parameters.
Although we are unaware of a complete analytic solution to these equations (\ref{adiva}-\ref{adivvarpi}), numerical solutions can be used to explore the phase space and help us to understand the evolution of the binary's orbit.

\subsection{Equations of motion}

As the majority of observed polluted white dwarfs are nearby \citep[\eg within  0.1kpc ][]{Holberg}, we restrict ourselves to the Solar neighbourhood, which we assume resides within 8kpc of the Galactic centre. This restriction allows us to neglect planar Galactic tides
(see Fig. 2 of \citealt*{VerasEvans2013MNRAS}), which represents a widely-used
simplification~\citep{Heisler1986,Matese1989,Matese1992,Matese1995,Breiter1996,Brasser2001,Breiter2005}.  Further, we consider only bound
binary systems.  Boundedness effectively equates to adiabaticity in the Galactic tidal
regime because no escape can occur in the adibatic regime \citep{VerasEvans2013MNRAS} and the
non-adibatic limit is close to the system's Hill ellipsoid \citep{Veras2014}. For very wide binaries ($a_{b}>>10^5$AU) adiabicity is no longer a good approximation (shown by Fig. 3 of \cite{VerasEvans2013CM}) and the full equations of motion describing this regime are shown in Column 5 of Table 1 of \cite{Veras2014}.

Here, we consider the adiabatic regime. Consequently we can, by considering vertical tides alone and using the perturbed two-body problem, derive the equations of motion for the orbital elements of the binary (as shown in \cite{VerasEvans2013MNRAS}); semi-major axis, $a_b$,  mean motion, $n_b$, eccentricity, $e_b$, inclination of the binary orbital plane to the Galactic plane, $i_b$, argument of pericentre, $\omega_b$ :

\begin{eqnarray}
\frac{dn_b}{dt}
&=& 0
\label{adiva}
\\
n_b\frac{de_b}{dt}
&=&
-\frac{5e_b \sqrt{1 - e_b^2}}{2} \cos{\omega_b} \sin{\omega_b} \sin^2{i_b}
\Upsilon_{zz}
\label{adive}
\\
n_b\frac{di_b}{dt}
&=&
\frac{5e_b^2\sin{2\omega_b}\sin{2i_b}}{8\sqrt{1 - e_b^2}}
\Upsilon_{zz}
\label{adivi}
\\
%\left(\frac{d\Omega}{dt}\right)_v
%&=&
%\frac{\cos{i} \left(2 + 3 e^2 - 5 e^2 \cos{2 \omega} \right)}{4n\sqrt{1 - e^2}}
%\Upsilon_{zz}
%\label{adOvi}
%\\
n_b\frac{d\omega_b}{dt}
&=&
\frac{5\sin^2{\omega_b} \left(\sin^2i_b - e_b^2\right) - \left(1 - e_b^2\right)}{2\sqrt{1-e_b^2}}
\Upsilon_{zz}
\label{adivvarpi}
\end{eqnarray}

The variable $\Upsilon_{zz}$ is determined by the Galactic model used, and contains information
about the matter density and gravitational potential.  Here, we use the disc
component of the three-component model of \cite{VerasEvans2013}.  Because we fix the distance from
the Galactic centre at $8$ kpc, we also fix $\Upsilon_{zz} = -5.352 \times 10^{-30} {\rm s}^{-2}$.

\subsection{An example binary}
\label{sec:example}
Although we are unaware of a complete analytic solution to equations (\ref{adiva}-\ref{adivvarpi}), we
can obtain numerical solutions that enable us to follow the evolution of example orbits. There are several key properties of the evolution of binary orbits under the influence of the Galactic tide. Of particular relevance to this work is the evolution of the binary's semi-major axis and eccentricity. The semi-major axis remains constant (Eq.~\ref{adiva}), under the adiabatic approximation of these equations. %{\bf (mention and explain the small changes seen by Nate? } . 
The eccentricity evolves periodically, with a time period and amplitude that depend on the initial conditions of the binary's orbit, as well as the strength of the Galactic tide ($\Upsilon_{zz}$). If a binary's orbit is to perturb the primary's planetary system, we are interested in the maximum eccentricity that its orbit reaches and the times between eccentricity peaks.  Thus, we focus on these two parameters in the following sections.

\begin{figure}
\includegraphics[width=0.48\textwidth]{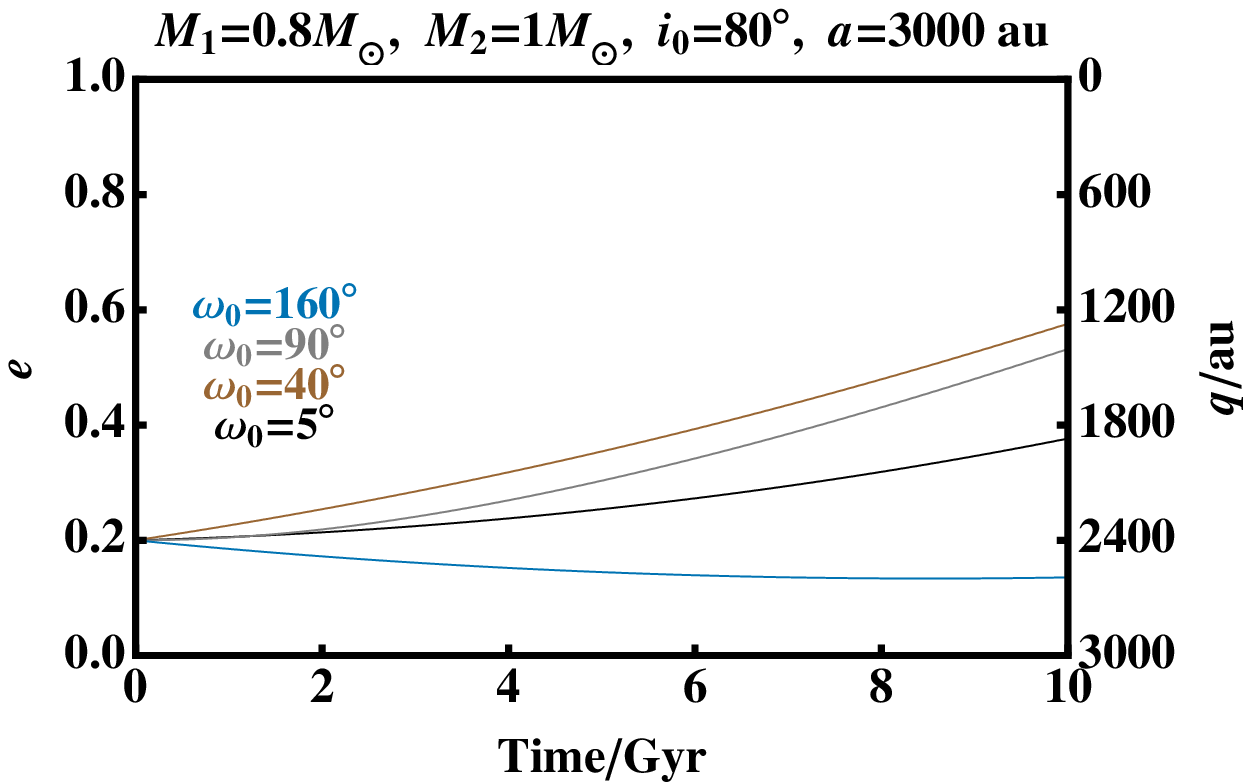}
             \includegraphics[width=0.48\textwidth]{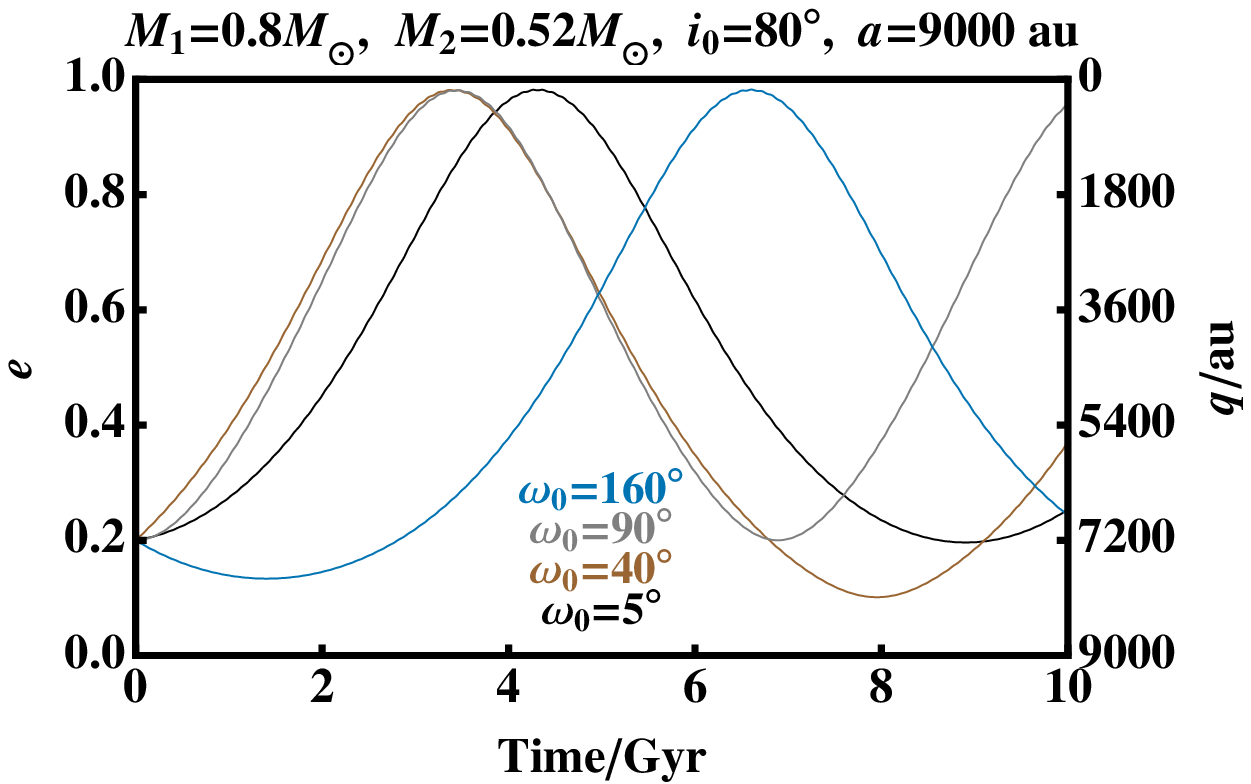}
\caption{
The change in orbital parameters due to Galactic tides and stellar evolution, for several example binaries. The top panel represents the primary on the main-sequence and the bottom panel, once the primary has evolved to become a white dwarf. Tides can produce extreme eccentricity oscillations during the white dwarf phase which do not occur along the main sequence, thereby triggering planetary scattering which can ultimately lead to white dwarf pollution.  Assumed in these plots is that the binary orbit expands non-adibatically by a factor of 3 due to mass loss from the secondary star $M_2$.
}
\label{fig:example}
\end{figure}

Firstly, in order to illustrate our proposed scenario, let us consider, for example, a $1M_{\odot}$ binary companion orbiting a $1M_\odot$ primary, in the solar neighbourhood ($R=8$kpc), with a semi-major axis of 3000AU, inclined to the Galactic plane by $80^\circ$. We gave the binary an initial eccentricity of $e_0=0.2$, and a range of initial arguments of pericentre ($\omega_0=5,40,90,160^\circ$) and followed the evolution of the binary's eccentricity, shown in the top panel Fig.~\ref{fig:example}. The eccentricity of this binary's orbit evolves on time periods that are significantly longer than the main-sequence lifetimes of the stars, or even the age of the Universe. Therefore, if the binary starts at low eccentricity, as we assume, it remains at low eccentricity throughout the primary's main-sequence lifetime.

We then consider the fate of this binary system, as the primary becomes a white dwarf. So far, the primary's planetary system is relatively unperturbed by the binary companion. By using a typical mass loss prescription\footnote{The {\tt SSE} code \citep{sse} applies Reimer's mass loss \citep{reimers78} at early evolutionary stages, but during the AGB it applies the semi-empirical mass-loss rate of \cite{VWood}, reaching a maximum during the superwind of $1.36\times 10^{−9} (L_* /L_\odot) M_\odot {\rm yr}^{-1}$, where $L_*$ is the stellar luminosity.}, the primary becomes a white dwarf of mass $0.52M_\odot$ (see caption of Fig.4 of \cite{Veras_twoplanet_2013}). At this semi-major axis  mass loss is likely to be non-adiabatic (see for example Fig.3 of \citep{Veras2014}), thus the exact expansion of the binary's orbit is difficult to predict \citep{Veras11}. We, therefore, make a reasonable assumption that the binary's orbit expands by a factor of 3 to $a_b=9,000$AU. Both the reduction in mass of the binary and the increase in semi-major axis of the binary, increase the ability of Galactic tides to change the binary's orbit. As is shown in the bottom panel of Fig.~\ref{fig:example}, the binary's eccentricity now evolves on `shorter' timescales, several rather than tens of Gyrs. Thus, with our assumption that the binary's orbit starts the white dwarf phase with an eccentricity of $e_0=0.2$, depending on $\omega_0$, it evolves to its maximum eccentricity (minimum pericentre) within 3-7Gyrs. Approaching the primary at a distance of closest approach of around 150AU, the binary is likely to perturb the primary's planetary system. It is these perturbations that we invoke for their ability to produce polluted white dwarfs, particularly at late times.

\subsubsection{Including mass loss}

\begin{figure}
\includegraphics[width=0.48\textwidth]{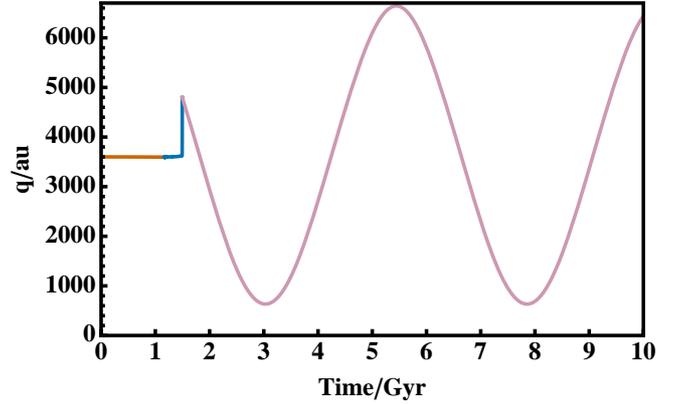}
\caption{A demonstration of how the first close approach in a binary system can occur during the white dwarf phase.  Plotted is the self-consistent pericentre distance evolution of one particular binary system through all phases of stellar evolution.   The initial binary parameters are:  $M_1 = 2 M_{\odot}$, $M_2 = 1 M_{\odot}$, $e_b = 0.1$, $a_b = 4000$ au, $i_b = 110^{\circ}$, $ \omega_b = 0^{\circ}$.  The red, blue and purple curves represent the main sequence, giant branch, and white dwarf phases of evolution, respectively.  Adiabatic Galactic tidal equations are used for the main sequence and white dwarf phases; the full non-adiabatic equations for both tides and mass loss are used along giant branch phases, with $f_0 = 150^{\circ}$.  }
\label{fig:fullevol}
\end{figure}

Fig.~\ref{fig:example} integrates Eqs.~\ref{adiva}-\ref{adivvarpi} for the main-sequence and the white dwarf phases separately. It is possible to integrate these equations self-consistently, if a formulation for stellar mass loss is included. However, it is computationally intensive and therefore, we only perform this calculation for one example system. Using the stellar evolution prescription of \cite{sse}, coupled with N-body integrations, as in \cite{Veras11}, Fig.~\ref{fig:fullevol} shows the evolution of the pericentre of a binary orbit that starts with a $2M_\odot$ and a $1M_\odot$ planet on an orbit with $a_b=4,000$AU, but where the $2M_\odot$ primary evolves to become a white dwarf of mass 0.64M$_\odot$. The same behaviour as in Fig.~\ref{fig:example} is seen. The binary's orbit is not perturbed significantly during the main-sequence evolution, but undergoes long period oscillations once the star becomes a white dwarf.

The mass loss is modelled to be isotropic, which represents an excellent approximation to reality \citep{VerasHadjTout2013}. In this case, the pericentre can never decrease during giant branch mass loss (Eq.~21 of \cite{Veras11}).

\section{The pollution of the white dwarf}
\label{sec:pollution}

We have shown that Galactic tides can perturb orbits such that a secondary companion approaches close to any planetary system orbiting the primary (white dwarf).  In this section we present simulations that illustrate how the close approach of the secondary could perturb the planetary system orbiting the white dwarf in such a manner that planetesimals are scattered onto star-grazing orbits.  For the purpose of this work we consider such scattering sufficient to produce white dwarf pollution, see \S\ref{sec:discussion} for a discussion of this assumption.  These simulations are intended as a proof of concept, rather than a detailed analysis.  

\subsection{Details of our simulations}
\label{sec:simulations}

Clearly, the parameter space available to investigate, both in terms of the diversity of planetary system architectures and binary orbits, is enormous. We can only investigate example systems. We, therefore, consider a binary in which a $1M_\odot$ primary is orbited by a companion of mass $0.8M_\odot$. The $1M_\odot$ primary evolves to become a $0.52M_\odot$ white dwarf \citep{Veras_twoplanet_2013}. $0.8M_\odot$ represents a typical value for a main-sequence stellar mass \citep{Parravano2011}, and such stars will remain on the main-sequence for the current age of the Universe ($\sim14$ Gyrs).

The primary is orbited by a very simplistic planetary system with one planet and one planetesimal belt or debris disc, similar to Neptune and the Kuiper belt in our Solar System. The presence of both the planet and planetesimals means that perturbations to the orbit of the planet can lead to planetesimals being scattered onto star-grazing orbits \citep{bonsor11, debesasteroidbelt}. In addition, the binary companion can perturb planetesimals directly onto star-grazing orbits \citep[e.g.][]{Marzari2005}.  We fix the plane of the planetary and binary orbits to be the same. We fix the planet's semi-major axis at 30AU, which mimics Neptune's semi-major axis, and the planetesimal belt runs from 30-50AU. We assume that at such a small orbital distance stellar mass loss is adiabatic, such that the white dwarf is orbited by a planet at 60AU and belt running from 60-100AU.  We neglect the potentially destructive effects of YORP spin up on the planetesimals \citep{VerasYORP} and the potentially significant movement of these asteroids due to the Yarkovsky effect from giant branch stellar evolution \citep{VerasYarkovsky}.  These effects would have a greater influence on an exo-asteroid belt located within 10AU than an exo-Kuiper belt located at several tens of AU.

Our simulations are run using the {\tt Mercury} N-body integrator, RADAU \citep{chambers99}. We track particles that are scattered close to the star.  It is these particles that we assume are tidally disrupted and pollute the white dwarf. The RADAU integrator is used in order to track the evolution of these particles to small pericentre, in a manner that the sympletic integrator cannot.  However, in order to limit the runtime for our simulations, we only track particles as close in as 0.1AU. We consider the particles that are scattered interior to 0.1AU to be representative of the population scattered onto star-grazing orbits where they would be disrupted, more likely to occur when $r<\sim R_\odot$ \citep{Veras_tidaldisruption1}.  Given that our purpose is to illustrate the feasibility of this mechanism, we consider this assumption to be sufficient. Our choice merely means that we over-estimate the population scattered onto star-grazing orbits, but do so in a consistent manner in all our simulations, thereby removing comparison inconsistencies between simulations.

We use our simulations as a tool to investigate the binary orbits that perturb the planetary system and scatter particles onto star-grazing orbits. Therefore, we focus on short integrations (10-100 binary orbits) for which the binary orbit is fixed and does not evolve due to the Galactic tide. These integrations can represent the evolution of the planetary system directly following the binary's evolution onto the orbit considered.

The exact orbits of the planet and planetesimals will clearly evolve with time and the evolution of the binary's orbit. It is not possible to consider all possible evolutionary paths. Instead we investigate the increase in the scattering of planetesimals onto star-grazing orbits in comparison with a simulation in which the binary is not present. Although this comparison may not always mimic the potential effects of an excited planetesimal belt, or scattered disc, it illustrates clearly the perturbative effect of the binary. We start our planetesimals with orbits that are initially unperturbed by the planet and have semi-major axes, eccentricities, inclinations, longitudes of pericentre, longitudes of ascending nodes and mean anomalies, which are randomly drawn from uniform distributions with ranges of $60<a<100$AU, $0<e<0.02$, $0<i<1^\circ$ and $0^\circ< \lambda, \Omega, M <360^\circ$.

\begin{figure}

\includegraphics[width=0.48\textwidth]{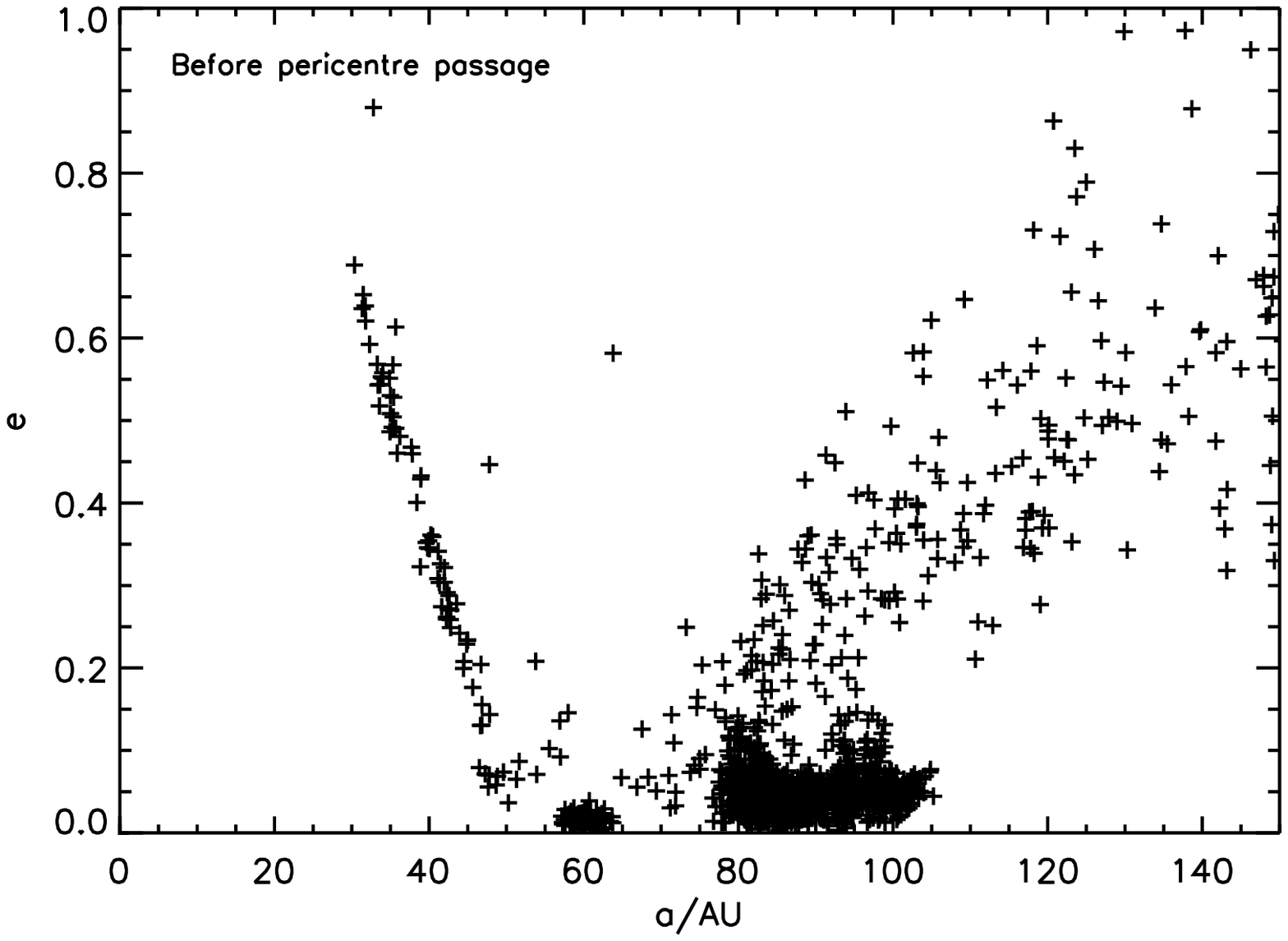}
\includegraphics[width=0.48\textwidth]{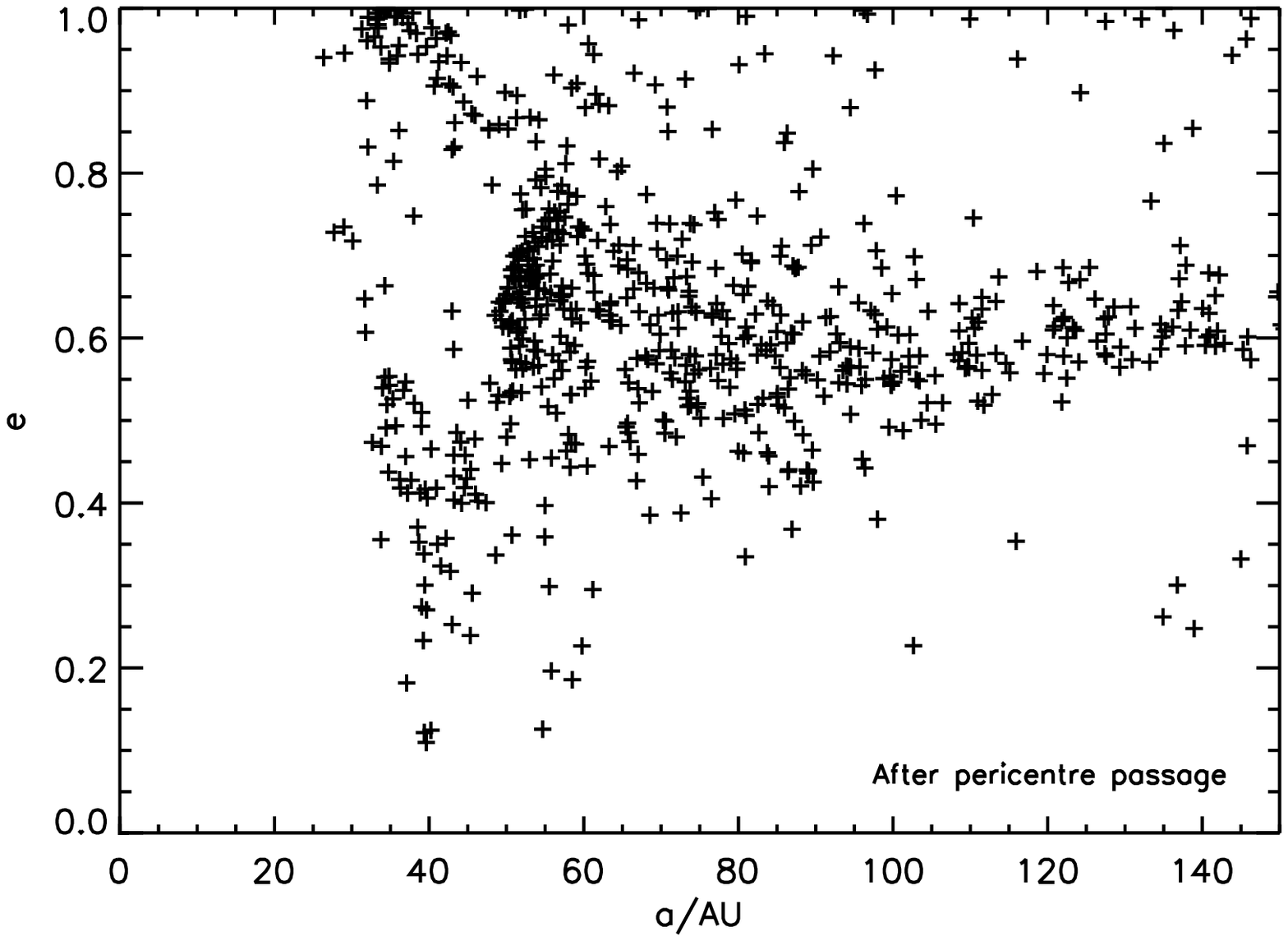}

\caption{ The semi-major axis, eccentricity distribution of planetesimals orbiting the white dwarf, prior and following the pericentre passage of a wide binary companion with $e_b=0.99$. This is the same system as shown in the bottom panel of Fig.~\ref{fig:example}.   This figure demonstrates that despite being dynamically excited, an exo-Kuiper belt largely survives stellar evolution of a primary star with a wide-orbit companion.}
\label{fig:ab9000}
\end{figure}
\begin{figure}

\includegraphics[width=0.48\textwidth]{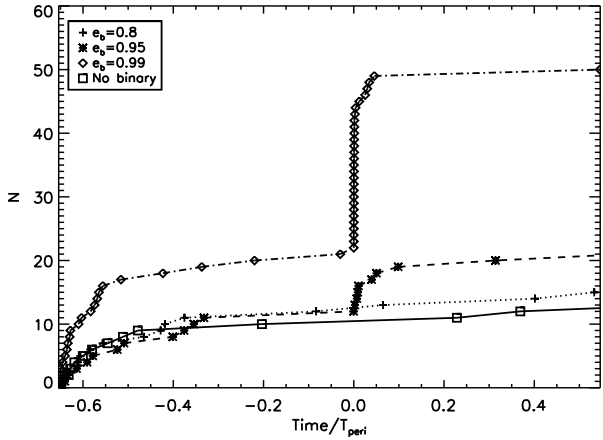}

\caption{ Our example white dwarf planetary system, orbited by a binary companion with $a_b=9,000$AU (shown in Fig.~\ref{fig:example} and Fig.~\ref{fig:ab9000}). The number of particles (N) that hit the star, as a function of time, in units of the binary's orbit period, with $t=0$, occurring at pericentre. The high binary eccentricity induced by the Galactic tide increase the number of planetesimal collisions with the white dwarf.}
\label{fig:ab9000_hitsun}
\end{figure}

\subsection{An example system}
\label{sec:example_polluted}

The aim of this section is to illustrate the manner in which the change in orbit of a binary due to Galactic tides can lead to the pollution of a white dwarf, even a white dwarf with a cooling age of Gyrs or longer. We consider the example binary system shown in Fig~\ref{fig:example}. The binary companion orbits at $a_b=3,000$AU and whilst it is on the main-sequence has its eccentricity perturbed very little by the Galactic tide, up to an absolute maximum of $e_b=0.6$, from $e_b=0.2$ initially, during a long main-sequence lifetime of 10Gyr. 
Even if the binary remains on an orbit with $e_b=0.6$ for 900Myr, a typical main-sequence lifetime, an outer belt survives orbiting the primary star. In fact, our test simulation finds that more than 60\% of the original belt remain on stable orbits. Given that Fig.~\ref{fig:example} shows that the binary's orbit evolves and will only be on such an eccentric orbit for timescales significantly shorter than the primary's full main-sequence lifetime. Thus, in this example, we show that a planetary system can survive the star's main-sequence lifetime without being perturbed unduly by its binary companion.

\begin{figure}
\includegraphics[width=0.45\textwidth]{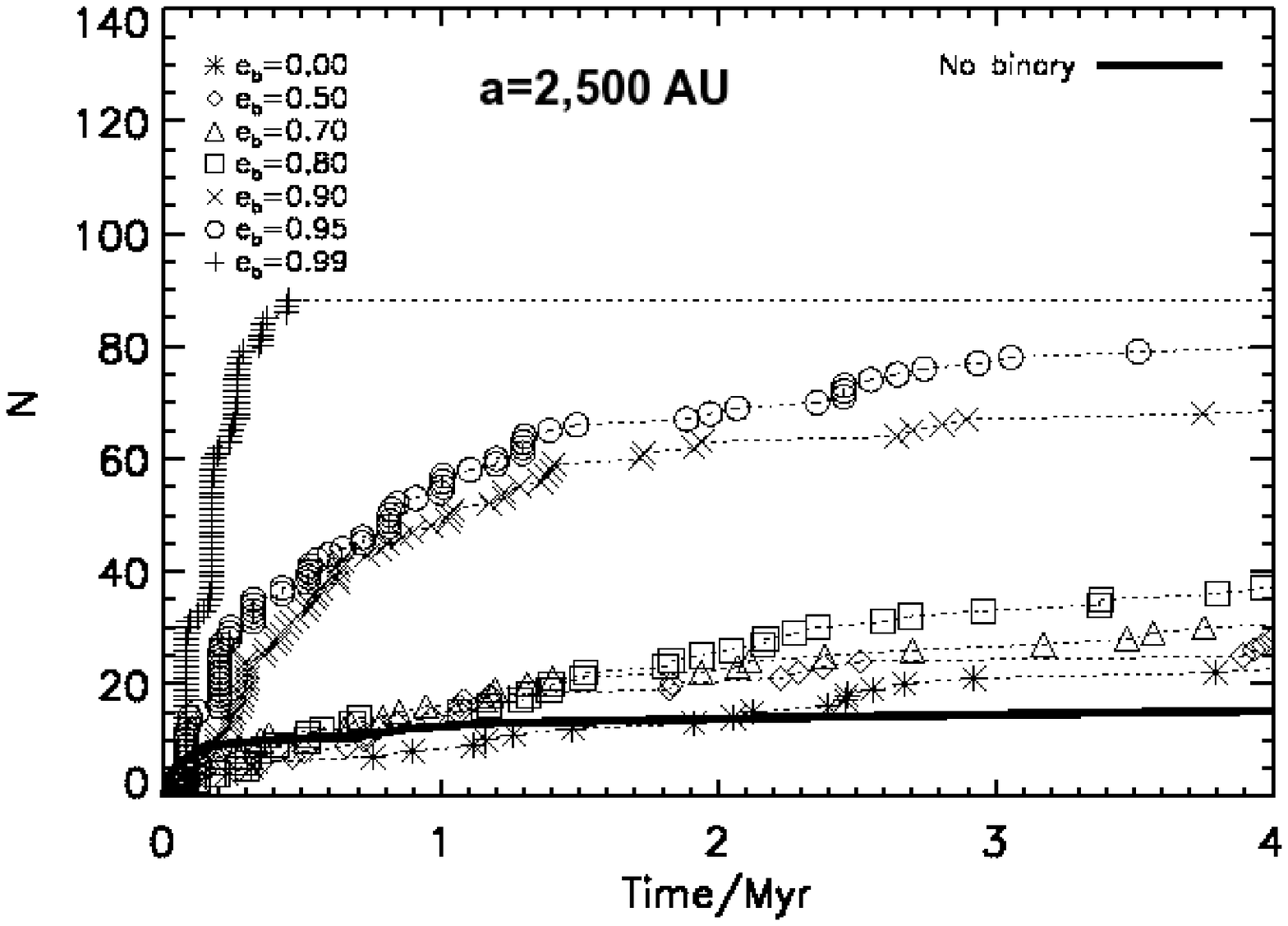}
\includegraphics[width=0.45\textwidth]{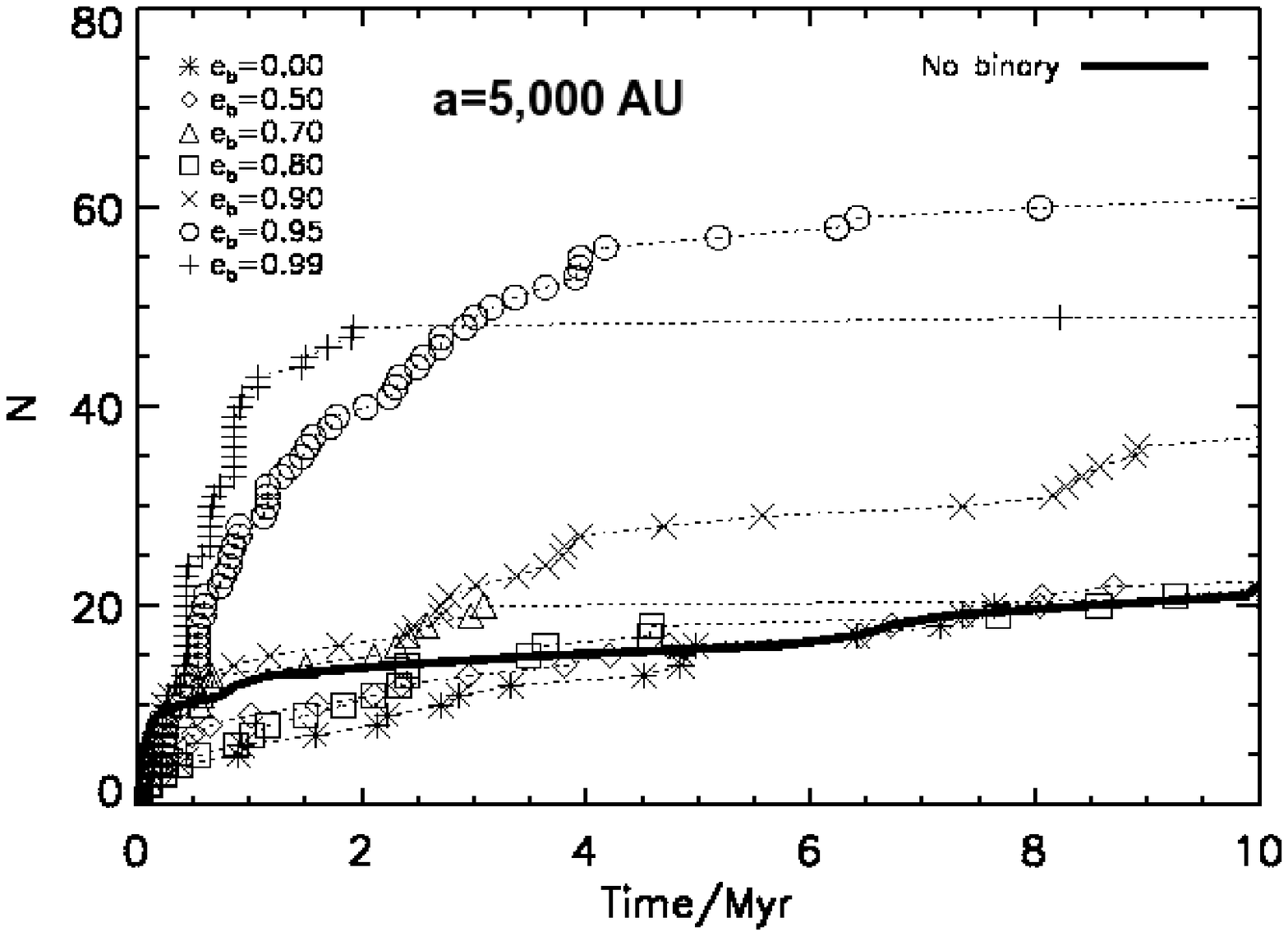}
\includegraphics[width=0.45\textwidth]{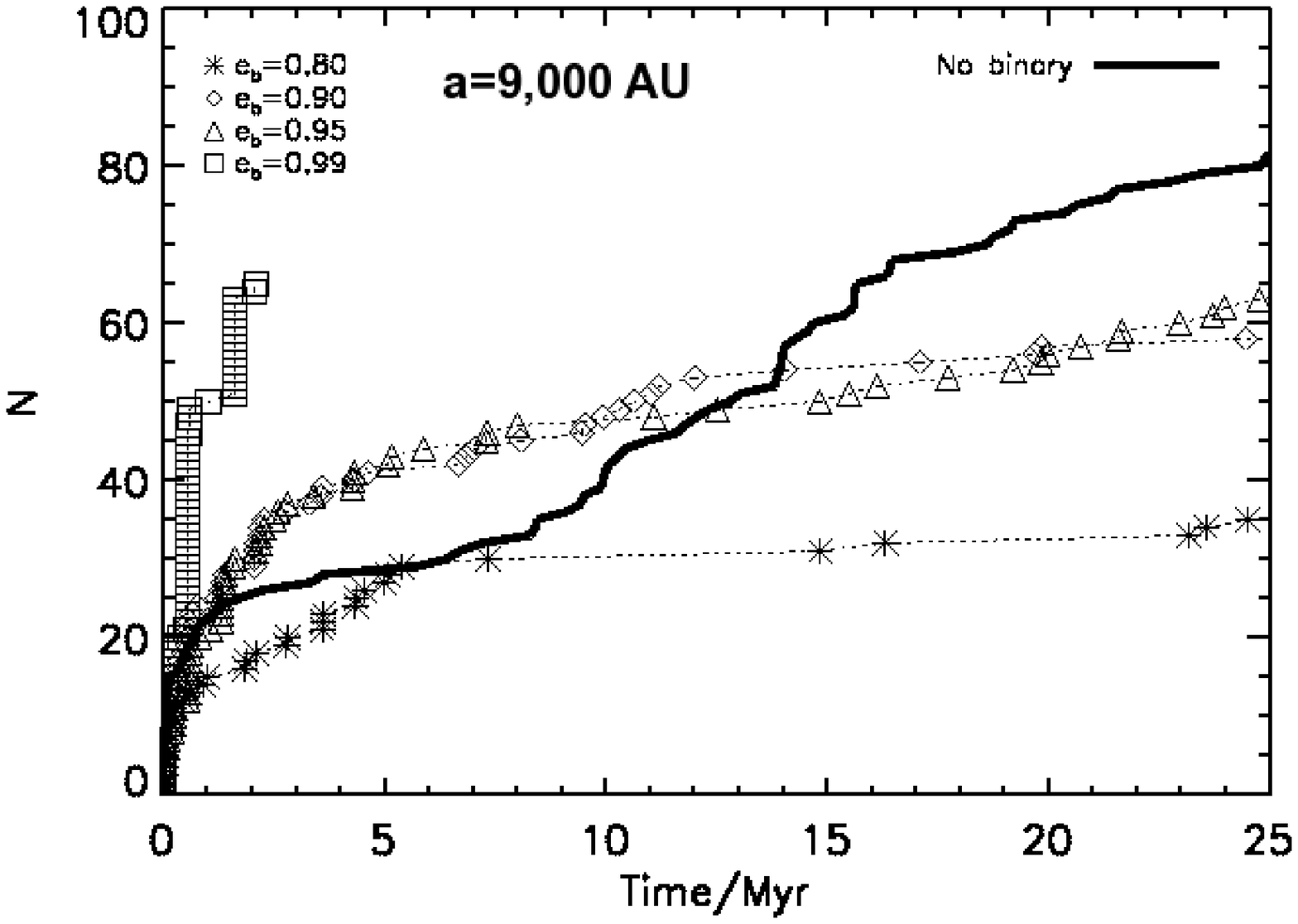}

\caption{ The number of particles (N) scattered onto star-grazing orbits as a function of time, for $a_b=2,500$AU (top), $a_b=5,000$AU (middle) and $a_b=9,000$AU (bottom). The planet is ejected for $e_b=0.99$, $a_b=9,000$AU after 0.5Myr, hence the scattering stops.  This figure demonstrates that as the binary's orbit evolves to high eccentricity, the rate at which particles are scattered inwards increases significantly.  }% NB $a_b=9,000$AU has 2,000 particles not 1,000!}
\label{fig:hitsun}
\end{figure}

Once the primary star in this example binary evolves to become a white dwarf, the orbits of the binary and the primary's planetary system expand, {\it for the planetary system adiabatically, but for the binary non-adiabatically}. As above, we estimate that the binary's orbit moves to $a_b=9,000$AU and consider the perturbations that it now makes on the primary's planetary system, that has expanded adiabatically by a factor of 2 (belt is now at about 60-100AU). The binary's orbit is now perturbed by the Galactic tides, evolving up to an eccentricity of 0.978 ($q\sim200$AU) after several Gyrs of evolution on the white dwarf branch. As the binary's eccentricity increases, so does its influence on the planetary system, particularly during pericentre passages. Fig.~\ref{fig:ab9000} shows the effect on our example planetary system of a binary with eccentricity, $e_b=0.99$. Fig.~\ref{fig:ab9000} shows that prior to the binary's pericentre passage, the planetary system is relatively undisturbed, but that following its pericentre passage the planet and many planetesimals have been scattered. As discussed above, the scattered planetesimals have the potential to pollute a white dwarf. Fig.~\ref{fig:ab9000_hitsun} shows the number of planetesimals scattered onto star-grazing orbits following a pericentre passage by a binary on orbits with various eccentricities. For binary eccentricities above 0.8, significantly more planetesimals are scattered onto star-grazing orbits than without the presence of the binary. If we refer back to Fig.~\ref{fig:example}, depending on the exact parameters of the binary's orbit, the binary's eccentricity could be increased from 0.2 to above 0.8 between 2 and 6 Gyrs after the start of the white dwarf phase. In this manner we illustrate a mechanism that could lead to the pollution of old white dwarfs.

\subsection{A brief investigation of the parameter space}
\label{sec:paramspace}
The above section illustrated one example system in which white dwarf pollution could occur in white dwarfs with cooling ages of Gyrs. Clearly, the parameter space of binary orbits is huge and this mechanism will not work for all of them. A detailed investigation of the entire parameter space is beyond the scope of this work, and not necessary to illustrate the potential of this mechanism. Instead, we perform a small additional suite of N-body simulations in order to show that there are many binary orbits for which this mechanism could work. We focus here on the white dwarf phase only.

\begin{figure}
\includegraphics[width=0.48\textwidth]{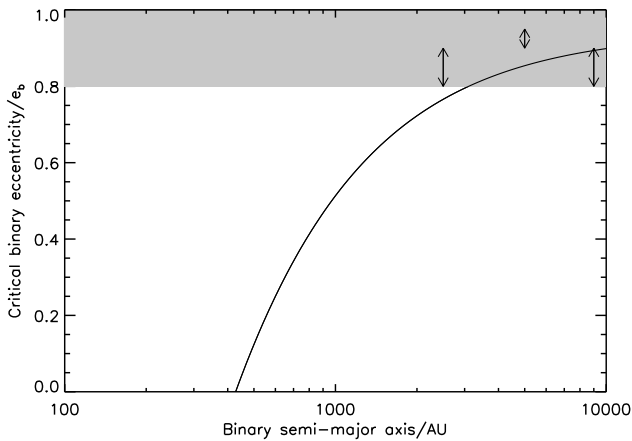}

\caption{The analytic criterion (solid line) for the stability of circumprimary orbits from \citet{Holman1999} (Eq. 1). The arrows show a comparison with the results of our simulations for the white dwarf phase, on short timescales. The shaded region indicates the limit of validity of this approximation, $e>0.8$, which unfortunately corresponds to the region of interest.   }
\label{fig:holman}
\end{figure}

We consider binary companions with semi-major axes of $a_b=2,500$AU, $a_b=5,000$AU and $a_b=9,000$AU. The eccentricity of the binary companion is varied, but fixed throughout each individual simulation.  Fig.~\ref{fig:hitsun} shows the number of planetesimals scattered onto star-grazing orbits as a function of time. The pericentre passages of the binary are evident from the increase in the scattering rate, creating sawtooth-like profiles. As noted in \S\ref{sec:pollution} these numbers must be compared to the simulation with the binary removed, in order to remove some dependencies on the initial conditions used. For sufficiently high eccentricity companions, it is clear that more material hits the sun than without a binary companion.  A conservative approximation to the eccentricity at which this occurs can be obtained from the criterion described in \cite{Holman1999} for the stability of planetary systems orbiting binary systems. Fixing $a_{crit}$ at 60AU in their Eq.1, although technically only valid for $e_b<0.8$, yields approximately the same critical eccentricities for the binary ($e_b>0.8$; $a_b=2,500$AU; $e_b>0.9$, $a_b=5,000$AU; $e_b>0.95$,$a_b=9,000$AU) as found from our simulations (see Fig.~\ref{fig:holman}).  Alternative criteria exist for the stability of three body systems \citep{Eggleton1995, Petrovich2015}, and although valid for higher eccentricity, have not been derived for two planets orbiting one star, rather than a binary companion and a planet orbiting the central star.

The critical point to take from these simulations is that planetesimals are scattered onto star-grazing orbits whilst a binary companion has high eccentricity, but remain on {\it more stable} orbits whilst the binary's eccentricity is lower.  Thus, planetesimals may survive the star's main-sequence evolution, as well as some proportion of its white dwarf evolution on stable orbits, before being scattered onto star-grazing orbits as the binary's orbit evolves to high eccentricity.

\subsection{Caveats and Assumptions}

The purpose of the above simulations is to illustrate that a wide binary companion could lead to white dwarf pollution at late times by scattering planetesimals onto star-grazing orbits.  This requires that sufficient material survives to the white dwarf phase, as well sufficient material is scattered onto star-grazing orbits to produce detectable pollution.

We assume here that planetesimals scattered onto star-grazing orbits lead to white dwarf pollution.  This is a fairly robust assumption, as discussed in previous work, although gaps still exist in our knowledge of the exact processes involved \citep[\eg][]{JuraWD03, bonsor11, debesasteroidbelt, Frewen2014}.  Planetesimals scattered close to the star are tidally disrupted, and accrete onto the star \citep{Veras_tidaldisruption1, Veras_tidaldisrupt2}, potentially via dusty or gaseous accretion discs \citep{Hartmann11, rafikov1, rafikov2, Bochkarev2011, Metzger2012, Rafikov2012}.   Our simulations are good at showing that material is scattered onto star-grazing orbits, but only indicate increased rates of scattering, rather than exact levels, and cannot, therefore, be used to predict accretion rates.  However, the amount of material required to produce  even the most polluted systems (\eg a Ceres mass $\sim 10^{-4}M_\oplus$ for \cite{Dufour10} or see Fig.~9 of \cite{Girven2012}) is low, and our simulations indicate high levels of scattering (Fig.~\ref{fig:hitsun}).  These masses are low compared to typical planetesimal belt masses thought to exist on the main-sequence, even those below detection limits \citep{wyatt07, wyattreview}, such as the Kuiper belt with its mass of around 0.1$M_\oplus$ \citep{Gladman2001}.  Although these masses may be reduced by collisions, or dynamics as material is scattered during the star's main-sequence lifetime, we do not anticipate that in most systems these processes remove sufficient material to prevent detectable pollution being produced \citep{bonsor10}, although of course in some planetary systems dynamical clearing may be highly efficient \citep[\eg][]{Raymond2009b, Raymond2010}.

We choose to only simulate binaries with fixed eccentricity, in order to simplify our results and save computational time.  From this choice we can clearly show an increased rate of scattering relative to simulations where the binary is not present.  However, we miss any changes due to the exact evolution of the system as the binary's orbit increases smoothly in eccentricity, or due to particularities of the exact configuration of the system (\eg planetesimals with excited eccentricities or inclinations or trapped in resonance \etc).  These details may change the scattering rates.  However, given our vague knowledge of the structure of any planetary system, we consider that the present simulations are sufficient to show that material is scattered and a white dwarf could be polluted in this manner.

\section{Estimating the frequency of pollution caused by a wide binary companion}
\label{sec:fraction}
Having shown that for particular orbital parameters of the binary and planetary system, pollution of the white dwarf can occur, it would be useful to estimate the contribution of this mechanism to the population of polluted white dwarfs.  This value firstly depends on the binary having an appropriate orbit that remains at low eccentricity for the primary's main-sequence evolution, but increases to sufficiently high eccentricity to perturb the primary's planetary system during the primary's white dwarf evolution, and secondly, on the primary having a planetary system with sufficient material that can be scattered onto star-grazing orbits in order to produce the pollution.

\begin{figure}
\includegraphics[width=0.48\textwidth]{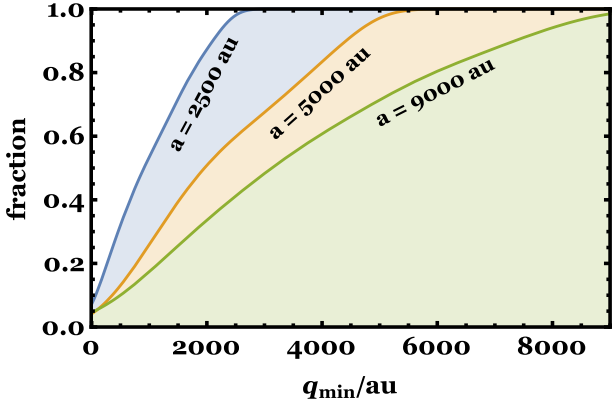}
\caption{ Estimating the fraction of systems where the wide binary mechanism has the potential to pollute the primary's planetary system. Plotted is the cumulative fraction of systems that have a pericentre passage closer than a critical value during the white dwarf phase.   The maximum eccentricity (minimum pericentre) can be estimated from Fig.~\ref{fig:hitsun}; see discussion in \S\ref{sec:binaryorbit} for details.   }
\label{fig:hist}
\end{figure}

\subsection{The binary orbit}
\label{sec:binaryorbit}
Firstly, we note that the binary orbits that are most
likely to be interesting in terms of the proposed scenario
are those at intermediate wide orbits (\ie $3, 000 \lesssim a_b \lesssim 10, 000$AU).  Here, the influence of Galactic tides is strong,
but not sufficiently strong that the main-sequence planetary
system would have been perturbed (as might occur for larger
semi-major axes) and the binary does not orbit too close to
the primary, so that the primary's planetary system is only
perturbed when the binary's orbit increases significantly in
eccentricity.

 Here, we estimate the fraction, $f_{\rm WB}$, of wide binaries where the eccentricity of the binary's orbit increases sufficiently (above a critical value), during the white dwarf phase, such that the primary's planetary system is perturbed.  
The distribution of orbital parameters of wide binaries in the galaxy are not well known. We, therefore, consider a uniform distribution of initial binary eccentricity, $e_0$ , longitude of pericentre, $\omega_0$, sine of inclination to
the Galactic plane, $\sin i_0$, and use a Monte-Carlo model to select 300 binary orbits for each semi-major axis\footnote{Tests found no significant difference betweeen using N=100 and N=300, so N=300 is assumed to be sufficient.}.  We fix the binary semi-major axis at $a_b=9,000$AU, $a_b=2,500$AU or $a_b=5,000$AU and the masses of the pair
fixed at $M_{W D} = 0.52M_\odot$ (evolved from $m_1 = 1M_\odot$ ) and $M_2 = 0.8M_\odot$), in line with Fig.~\ref{fig:example} and Fig.~\ref{fig:hitsun}. Fig.~\ref{fig:hist} shows the fraction of systems where a pericentre passage closer than a critical value (or equivalently, the eccentricity evolves to above a critical value), at any point during the white dwarf phase.  A maximum time period of 14Gyr, the age of the Universe, is considered.  From Fig.~\ref{fig:hitsun}, an eccentricity of greater than approximately 0.8 ($a_b=2,500$AU), 0.9($a_b=5,000$AU) or 0.95 ($a_b=9,000$AU) is required to perturb a planetary system at 60-100AU. In other words, the binary must enter the region interior to $\sim 500$AU to perturb the planetary system.  Fig.~\ref{fig:hist} shows that 35\%, 17\% and 10\% of systems evolve to this pericentre.  Although clearly this value will vary significantly with the binary's semi-major axis, we consider 20\% to be a reasonable estimate for the fraction of wide binaries with orbital parameters in the correct range to produce polluted white dwarfs.

Another constraint on the orbit of the binary results from the requirement that a planetary system survives, relatively unperturbed, orbiting the primary until late times. This requirement means that the binary's eccentricity must remain below a critical value during the primary's main-sequence evolution.  This value can be roughly estimated by considering the criterion of \cite{Holman1999} (Eq 1.) for stable circumprimary orbits. For example for our three example binaries with $a_b=2,500$AU, 5,000AU and 9,000AU, the binary eccentricity must remain below $e_b<0.4,0.67$ and 0.8 (assuming $a_b\sim800$AU, 1,600AU and 3,000AU on the main-sequence), which is not a very strict constraint given the reduced perturbations due to the Galactic tide whilst the primary is on the main-sequence and the binary orbit tighter.  Given the other uncertainities in our estimation of the fraction of white dwarfs polluted by the wide binary mechanism, we leave this constraint out of our estimation.

\subsection{The planetary system}
If $f_{\rm WB}\sim 20\%$ of wide binaries have orbital parameters such that a close approach occurs during the primary's white dwarf evolution, the next question regards what fraction, $f_{PS}$, of these white dwarfs have planetary systems that can be scattered in such a manner as to produce white dwarf pollution.  Our simulations (\S\ref{sec:pollution}) necessarily consider a very specific architecture for the planetary system. Scattering, however, would occur in a wide range of planetary system architectures, including Oort cloud-like structures, rather than Kuiper-like belts.   Clearly, the architecture of the planetary system has a great influence on the efficiency at which material is scattered inwards \citep[\eg][]{bonsor_exozodi}, however, for the purposes of this estimation we consider it sufficient to assume that all planetary systems where sufficient material survives have the potential to produce white dwarf pollution via the wide binary mechanism.  An advantage of this mechanism is its lack of dependence on the exact structure of the planetary system.  Outer planetary systems should survive the star's evolution to the white dwarf phase, although planetesimal belts may become collisionally depleted \citep{bonsor10}, and dynamical instabilities may clear some systems of material \citep[\eg][]{Raymond2009b}.  Given the prevalence of planetary systems in the Solar neighbourhood \citep[\eg][]{Borucki2011, Mullally2015} and that most planets or planetesimal belts cannot be detected, we consider a reasonable estimate to the fraction of wide binaries, where the white dwarf primary has a planetary system suitable for producing pollution to be $f_{PS} \sim 50\%$.  We note that this estimate is vague, and is actually something that observations of polluted white dwarfs may help to better constrain.

\subsection{Conclusion}

We can estimate the fraction of white dwarfs with wide binary companions likely to be polluted by this mechanism,  at some point during their lives as white dwarfs, as
\begin{equation}
f_{poll} \sim f_{\rm WB} \times f_{PS} \; \sim 20\% \times 50\% \sim 10\% 
\end{equation}
with large uncertainties on our estimations of all of the parameters. 
 At any given instance in time pollution would be detectable in only a fraction, $f_T$, of these systems.  This fraction will depend on the timescale on which pollution is detectable following a close approach of the binary star.  Many factors, of which are understanding is limited, need to be considered to produce a good estimate for this timescale, including the dynamical timescales in the planetary system, lifetime of any accretion disc formed, sinking timescale of metal pollutants, \etc.   We anticipate that the dynamical timescales in the planetary system will dominate, being several orders of magnitude longer than the other timescales involved here.  The dynamical timescales on which the outer planetary system continues to feed the white dwarf pollution may vary significantly depending on the exact architecture of individual planetary systems.  Rather than attempt to approximate these, we, therefore, conclude that a maximum of a few percent of any sample of white dwarfs with wide binary companions would be observed to have metal pollution produced via the wide binary mechanism presented in this work.  The wide binary fraction of white dwarfs is not well known, but if we consider that at least (and probably more than) 10\% of white dwarfs should have wide binary companions, based on the 10\% of solar-type stars observed with wide binaries \citep{Raghavan2010,LepineBongiorno2007} (see discussion in the Introduction), then the wide binary pollution mechanism contributes to pollution in a maximum of a percent of any sample of white dwarfs.

\section{Observational tests}
\label{sec:observations}

The mechanism presented here works only for a subset of the potential orbital parameters of the binary and planetary system.  As discussed in \S\ref{sec:fraction}, this fraction is likely to be of the order of a few percent of any random sample of white dwarfs with wide binary companions.  Given that many observations find that more than 25\%, and maybe up to 50\%, of white dwarfs are polluted \citep{Zuckerman03, ZK10, Koester2014}, this wide binary mechanism is only ever going to make a minor contribution, and other mechanisms must be important in polluting white dwarfs.  In fact, of roughly one hundred `pre-Sloan' polluted white dwarfs, discovered by their proper motions, such that any companions is likely to have been found, only 5 have known common proper motion companions (Farihi et al, private communication).  It is not clear whether this wide binary fraction is different from the general population, where, for example $\sim10\%$ of solar-type stars are found to have wide separation ($>1,000$AU) companions  \citep{Raghavan2010,LepineBongiorno2007}.

\cite{Zuckermanbinary} consider a sample of 17 white dwarfs with wide ($>2,500$AU) companions, all observed with Keck/VLT to search for Ca II, and find that 5/17 (30\%) are polluted.  With such small number statistics, and significant bias in sample selection, it is difficult to make a good comparison with observations of apparently single white dwarfs, where, for example 25\% of DA white dwarfs exhibit Ca pollution \citep{Zuckerman03}.  It remains plausible that the wide binary mechanism could contribute to the pollution in these white dwarfs on the percent level, but without larger, well defined, samples, it is impossible to make a detailed comparison.

The key evidence for a time-independent mechanism for pollution, such as suggested here, comes from the lack of any decrease in the fraction of polluted systems with age, and the observations of pollution in old (Gyr) white dwarfs, which, for example dynamical instabilities following stellar mass loss struggle to explain \citep[\eg][]{bonsor11, debesasteroidbelt, Frewen2014}.  There are currently insufficient white dwarfs with known binary companions to assess any dependence in the fraction of polluted systems with white dwarf cooling age.  However, WD 1009-184, with its companion at 6,870AU and an effective temperature of 9,940K (equivalent to a cooling age of $\sim$700Myr) \citep{wdcat09,  Zuckermanbinary} stands out as a candidate example of where the companion could potentially be responsible for the white dwarf pollution.  Future observational searches for companions to cool white dwarfs will provide critical evidence regarding the contribution of wide binaries to pollution.  Gaia will play a crucial role in discovering companions to many nearby white dwarfs.

\section{Discussion}
\label{sec:discussion}
In this work we outline a proof of concept that illustrates the manner by which a wide binary companion could lead to pollution in white dwarfs.  In \S\ref{sec:galatic} we illustrate that during the evolution of the binary's orbit as it interacts with the Galactic tides, periods of increased eccentricity occur.  These periods are most likely to occur during the white dwarf phase, due to the increase in separation of the binary orbit following stellar mass loss.  We claim that a wide binary companion whose eccentricity increases for the first time during the primary star's white dwarf evolution has the potential to perturb any planetary system orbiting the white dwarf.  In \S\ref{sec:pollution}, we illustrated how planetesimals can be scattered onto star-grazing orbits, a pre-requisite for pollution of the white dwarf's atmosphere.  This mechanism is independent of white dwarf cooling age, and could, therefore, produce pollution around old (1-5Gyr) white dwarfs that other mechanisms struggle to explain.  We estimate that this mechanism may contribute to the total fraction of polluted white dwarfs on the level of a few percent. There are no current observational samples that are sufficiently large to assess whether the population of polluted white dwarfs around single stars is different to those with companions.

We consider that this mechanism is robust and will occur for white dwarfs with planetary systems and companions with the `correct' parameters.  The weakest part of this argument regards the link between planetesimals scattered inwards and the pollution of white dwarf atmospheres, which was not investigated in this work.  The tidal disruption of planetesimals and the acccretion of close-in dusty material onto white dwarfs have been previously investigated \citep[\eg][]{rafikov1, rafikov2, debesasteroidbelt, Metzger2012, Veras_tidaldisruption1, Veras_tidaldisrupt2}, although some gaps still remain in our understanding.  Our discussion of whether a detectable level of pollution is produced is vague (see \S\ref{sec:fraction}), and it remains possible that no detectable pollution is ever produced by planetesimal scattering.  However, our simulations do show a significant increase in the number of planetesimals scattered onto star-grazing orbits with a binary companion on an eccentric orbit, compared to without a binary companion (see Fig.~\ref{fig:hitsun}), and given the low accretion masses required to produce detectable pollution\footnote{Even one of the most highly polluted white dwarfs only needs a Ceres mass ($10^{-4}M_\oplus$) \citep{Dufour10}, or see \cite{Girven2012}}, it remains plausible that the mechanism presented works for some region of the parameter space.  In fact, we anticipate that this mechanism has a weak dependence on the exact architecture of the planetary system, and a much stronger dependence on the orbital parameters of the binary.

The purpose of this work was to outline a potential mechanism for the pollution of white dwarfs.  We do not consider our estimation of the fraction of white dwarfs with wide binary companions polluted by this mechanism to be robust, and it is even trickier to determine the fraction of an observational sample that would be caught whilst displaying detectable pollution.  Our estimation does, however, indicate that the wide binary mechanism is not the dominant mechanism for producing white dwarf pollution, even at late times, and that statistics in current observational samples are insufficient to determine whether or not this mechanism is producing pollution.  Gaia and future observational searches for companions to nearby white dwarfs, should enable a clearer determination of its contribution.

We have ignored the potential for exchange interactions, or changes to the binary's orbit, which could have a big effect on the evolution of individual systems, but are not going to significantly affect the total population of white dwarfs with wide binary companions. We have also ignored any potential evolution of the companion star.  If it loses a significant fraction of its mass, this could increase the fraction of binary orbits that meet our criteria for producing pollution, as in essence the system has double the chance of entering the correct regime.  For example, one could imagine that the expansion in the orbit of the binary is insufficient to induce large perturbations from the Galactic tide as the primary loses mass to become a white dwarf, but becomes sufficiently wide once the secondary evolves to become a white dwarf.  Perturbations could occur to planetary systems orbiting either the primary or the secondary.

Our mechanism depends critically on the survival of a planetary system, containing sufficient material, to late times in the white dwarf phase.  Although, theoretically there are good arguments for the survival of outer planetary systems during post-main sequence evolution \citep{ DuncanLissauer, Veras_Wyatt2012}, even if collisionally depleted \citep{bonsor10}, there is a yet little direct observational evidence for the presence of planetary systems around white dwarfs \citep{Mullally08,dodo1,Xu2015}, although a few potential exceptions exist \citep{Luhman2011, Marsh2014}.  Planets have been detected around giant stars \citep[\eg][]{Johnson07,Johnson08, Sato08} and horizontal branch stars \citep{HBplanet}.  Dusty, debris-like discs have been observed at the centre of several planetary nebulae \citep{helix, Chu11}, very early in the white dwarf phase.  If future observations show that this mechanism is a significant contributor to white dwarf pollution, it provides further evidence for the survival of outer planetary systems to the white dwarf phase.

The most important contribution of the wide binary mechanism presented here is to the pollution of old (Gyr) white dwarfs, that are hard to explain by other mechanisms.  The mechanism is independent of white dwarf cooling age, except for any variation in the planetary system itself.  On the main-sequence material can be depleted, both by collisions and dynamics, with a steep time dependence.  By the white dwarf phase, collisional timescales are long \citep{bonsor10}, and the system is much more likely to have reached a dynamically stable state, such that any such evolution is likely to only occur at low levels.

\section{Conclusions}
\label{sec:conclusions}
In this work we present a mechanism to explain the metal pollution observed in many white dwarfs, especially at late times.  The orbits of wide binaries vary periodically with the Galactic tides.  The influence of the Galactic tide increases following stellar mass loss.  Thus, a wide binary may evolve for Gyrs at large separations, and yet, have a close approach for the {\it first time} during the primary's white dwarf phase.  Such a close approach could perturb any planetary system orbiting the primary, potentially scattering planetesimals onto star-grazing orbits.  

We present simulations that illustrate the change in orbital parameters of the binary (\S\ref{sec:galatic}) and the scattering of planetesimals onto star-grazing orbits (\S\ref{sec:pollution}).  These demonstrations should be considered a proof of concept, rather than a detailed analysis.  We estimate that this mechanism could contribute to pollution in up to a few percent of any sample of white dwarfs with wide binary companions, and up to a percent of all white dwarfs (\S\ref{sec:fraction}), independent of the age of the white dwarf.  Current observational samples do not have sufficient statistics to indicate whether or not this mechanism provides an important contribution to pollution in white dwarfs.  The calcium-polluted, WD 1009-184, with its companion at 6,870AU and an effective temperature of 9,940K (equivalent to a cooling age of $\sim$700Myr) \citep{wdcat09,  Zuckermanbinary} stands out as an example system where this mechanism could be acting.

Wide binary pollution is unlikely to be the sole contributor to the population of metal-polluted white dwarfs.  Instead, we envisage that it acts alongside other mechanisms, adding a time-independent pollution rate.  Future observational searches for wide binary companions to nearby white dwarfs (\eg with Gaia) and an increased sample of metal-encriched old, white dwarfs are required to determine whether or not wide binary companions play an important role in white dwarf pollution.

\section{Acknowledgements}
 We thank Silvia Catal\'{a}n and Jay Farihi for useful discussions.  We thank the referee Dr. Cristobal Petrovich for his useful comments that improved the quality of the manuscript. 
AB was supported by ANR-2010 BLAN-0505- 01(EXOZODI), NERC Grant NE/K004778/1, and ERC grant number
279973 whilst conducting this research. DV benefitted from support by the European Union through ERC Grant Number 320964.  
\bibliographystyle{mn}

\bibliography{ref}

%%%%%%%%%%%%%%%%%%%%%%%%%%%%%%%%%%%%%%%%%%%%%%%%%%

% Don't change these lines
\bsp	% typesetting comment
\label{lastpage}
\end{document}